\documentclass[aps,preprint,epsfig,rotate]{revtex4}
\begin{document}
\title{On the bound states in the muonic molecular ions}

 \author{Alexei M. Frolov}
 \email[E--mail address: ]{afrolov@uwo.ca}

\affiliation{Department of Chemistry\\
 University of Western Ontario, London, Ontario N6H 5B7, Canada}

\date{\today}

\begin{abstract}

The mass corrections to the bound state energies of the three-body muonic 
molecular ions $pp\mu, pd\mu, pt\mu, dd\mu, dt\mu$ and $tt\mu$ are 
determined numerically from the results of highly accurate computations. The 
total energies and some other bound state properties of these ions are 
evaluated to very high accuracy for the bound $S(L = 0)-, P(L = 1)-$ and 
$D(L = 2)-$states. In these highly accurate calculations we used the most 
recent and accurate masses of particles $m_p, m_d, m_t$ and $m_{\mu}$ known 
from high energy experiments. We also investigate some bound state properties
of the muonic molecular ions. In particular, we determine the hyperfine 
structure splittings of the ground states of the $pd\mu, pt\mu$ and $dt\mu$ 
ions. In these calculations we used our highly accurate expectation values of 
the interparticle delta-functions obtained in recent computations. The 
corresponding hyperfine structure splittings, e.g., $\Delta_{12} =$ 
1.3400149$\cdot 10^7$ $MHz$ and $\Delta_{23} =$ 3.3518984$\cdot 10^7$ 
$MHz$ for the $pt\mu$ ion, can directly be measured in modern 
experiments. Analogous hyperfine structure splittings are evaluated 
to very high accuracy for all five bound $S(L = 0)-$states in the three 
symmetric muonic molecular ions: $pp\mu, dd\mu$ and $tt\mu$. 

PACS number(s): 36.10.+Di, 36.10.-k and 31.10.+z.
\end{abstract}
\maketitle

\newpage

\section{Introduction}

In our earlier study \cite{FrWa2011} we considered the bound state spectra 
in the muonic molecular ions $pp\mu, pd\mu, pt\mu, dd\mu, dt\mu$ and 
$tt\mu$, where the notations $p, d, t$ stand for the nuclei of three 
hydrogen isotopes (protium, deuterium and tritium, respectively), 
while $\mu$ means the negatively charged muon $\mu^{-}$. In our calculations 
in \cite{FrWa2011} we have used the particle masses taken from relatively 
old experimental papers, since in \cite{FrWa2011} we wanted to show the 
progress achieved recently in highly accurate computations of Coulomb 
three-body systems with arbitrary masses. Therefore, it was some logic to use 
the same particle masses in all such calculations (see, e.g., \cite{Fro2001}, 
\cite{Fro01} and references therein). On the other hand, right now the masses 
of all nuclei of hydrogen isotopes ($p, d$ and $t$) and negatively charged 
muon $\mu^{-}$ are known to much better accuracy, than they were obtained ten 
years ago (in fact, eighteen years ago). It is clear that it is necessary to 
perform extensive recalculations of the bound state energies and other bound 
state properties in the six muonic molecular ions $pp\mu, pd\mu, pt\mu, dd\mu, 
dt\mu$ and $tt\mu$ by using the `recenty updated' particle masses.

Modern highly accurate computations of the bound states in muonic molecular 
ions allow one to determine 15 - 18 correct decimal digits in the total energy 
$E$. In some cases, e.g, for the $S(L = 0)-$state in the $pp\mu$ ion such an 
accuracy is much higher and we can determine $\approx$ 21 - 22 correct decimal 
digits in the total energy. On the other hand, the masses of particles have 
been determined to the accuracy which corresponds to $\approx$ 10 - 11 exact 
decimal digits only. Such uncertainties in particle masses lead to relatively 
large mistakes in the total energies and corresponding wave functions. Formally, 
this means an almost constant need of recalculation of the corresponding total 
energies and wave functions by solving the non-relativistic three-body 
Schr\"{o}dinger equation with the new masses. Note also that in contrast with 
the two-electron atoms for three-body muonic molecular ions we cannot use 
various mass-interpolation formulas for the total energies, since they are not 
very accurate and contain not one, but two and even three different parameters 
(i.e. the ratios of the masses of particles) and none of these parameters is 
small.

The main goal of this study is to perform the highly accurate computations of 
the bound states in the six muonic molecular ions $pp\mu, pd\mu, pt\mu, dd\mu, 
dt\mu$ and $tt\mu$. All particle masses used in our calculations are taken 
from the most recent high energy experiments. Our calculations are performed 
with the use of extended arithmetic precision. Finally, the mass corrections to 
the total energies of these ions have been determined (for each bound state in 
these six ions) to very high accuracy. 

By using the highly accurate expectation values of all (three) interparticle 
delta-functions obtained in our calculations we also investigate the hyperfine 
structure of the bound $S(L = 0)-$states in the six muonic molecular ions 
$pp\mu, pd\mu, pt\mu, dd\mu, dt\mu$ and $tt\mu$. The hyperfine structure 
splittings are determined for each of these (nine) bound $S(L = 0)-$states. The 
results of this investigation lead us to a number of interesting conclusions 
and observations. Many of these facts are important in analogous computations 
of more complicated systems, e.g., in the analysis of hyperfine structure 
splittings in the four-, five- and six-body quasi-atoms and ions which contain 
muonic molecular ions as a part of their structure.   

\section{The mass-dependent Hamiltonian of muonic molecular ions}

In the non-relativistic approximation the Hamiltonian of the three-body muonic 
molecular ion $a b \mu$ (or $(a b \mu)^{+}$) takes the form  
\begin{equation}
 H = -\frac{\hbar^{2}}{2 m_{\mu}} \Bigl( \frac{m_{\mu}}{m_a} \nabla^{2}_{a}
 + \frac{m_{\mu}}{m_b} \nabla^{2}_{b} + \nabla^{2}_{\mu} \Bigr) +
 \frac{q_a q_b e^2}{r_{ab}} + \frac{q_a q_{\mu} e^2}{r_{a{\mu}}} + \frac{q_b
 q_{\mu} e^2}{r_{b{\mu}}} \label{Ham}
\end{equation}
where $\nabla_{i} = \Bigl( \frac{\partial}{\partial x_{i}},
\frac{\partial}{\partial y_{i}}, \frac{\partial}{\partial z_{i}} \Bigr)$ and
$i = a, b, \mu$. In Eq.(\ref{Ham}) $\hbar$ is the reduced Planck constant 
($\hbar = \frac{h}{2 \pi}$) and $e$ is the elementary electric charge.  
It is very convenient (see below) to consider the bound state spectra of such 
ions in muon-atomic units in which $\hbar = 1, m_{\mu} = 1$ and $e = 1$. The 
speed of light $c$ in these units is $c = \alpha^{-1}$, where $\alpha = 
\frac{e^2}{\hbar c}$ is the fine structure constant. In muon-atomic units the 
same Hamiltonian, Eq.(\ref{Ham}), is written in the form
\begin{equation}
 H = -\frac12 \Bigl( \frac{1}{m_a} \nabla^{2}_{a} + \frac{1}{m_b}
 \nabla^{2}_{b} + \nabla^{2}_{\mu} \Bigr) + \frac{1}{r_{ab}} -
 \frac{1}{r_{a{\mu}}} - \frac{1}{r_{b{\mu}}} \label{Ham1}
\end{equation}
where the nuclear masses $m_a$ and $m_b$ of the two hydrogenic nuclei must be 
expressed in terms of the muon mass $m_{\mu}$. 
 
In our earlier studies (see, e.g., \cite{FrWa2011}, \cite{Fro2001} and 
references therein) we used the following masses of the hydrogenic nuclei 
$p, d, t$ and negatively charged muon $\mu$       
\begin{eqnarray}
 m_{\mu} = 206.768262 m_e \; \; \; , \; \; \; m_p = 1836.152701 m_e \\
 m_{d} = 3670.483014 m_e \; \; \; , \; \; \; m_t = 5496.92158 m_e \nonumber
\end{eqnarray}
where $m_e$ designates the electron mass. In particular, these masses were 
used in our earlier studies (see, e.g., \cite{Fro2001}, \cite{Fro01} and 
references therein). In this work the updated values for the nuclear masses of 
all four particles $p^{+}, d^{+}, t^{+}$ and $\mu^{-}$ will be used. The masses 
of these four particles have recently been determined in various high-energy 
experiments to better accuracy than they were known in the middle of 1990's. 
Usually, these masses are expressed in special high-energy mass units 
$MeV/c^{2}$. In these high-energy units the corresponding masses are 
\begin{eqnarray}
 m_{\mu} = 105.65836668(38) \; \; \; , \; \; \; m_p = 938.272046(21) \\
 m_{d} = 1875.612859(41) \; \; \; , \; \; \; m_t = 2808.290906(70) \nonumber
\end{eqnarray}
These values include current experimental uncertanties. They are very close to 
the masses of these particles given in \cite{CRC}. In all calculations 
performed in this study we have used the following masses of the $\mu, p, d$ 
and $t$ particles (in $MeV/c^{2}$)
\begin{eqnarray}
 m_{\mu} = 105.65836668 \; \; \; , \; \; \; m_p = 938.272046 \\
 m_{d} = 1875.612859 \; \; \; , \; \; \; m_t = 2808.290906 \nonumber
\end{eqnarray}
These values are considered as exact. The corresponding corrections to these 
masses can be taken into account by performing direct variational computations 
with the `new' masses. 
 
Since our calculations of muonic molecular ions are performed in muon-atomic 
units, then we need to use only three following mass ratios $\frac{m_{\mu}}{m_p}, 
\frac{m_{\mu}}{m_d}$ and $\frac{m_{\mu}}{m_t}$. These mass ratios are, in fact, 
the dimensionless parameters which determine the energy spectra and all other 
properties of the six muonic molecular ions $pp\mu, pd\mu, pt\mu, dd\mu, dt\mu$ 
and $tt\mu$. The total energies and the corresponding wave functions of the 
muonic molecular ions are determined during the highly accurate solution of the 
non-relativistic Schr\"odinger equation $H \Psi({\bf r}_1, {\bf r}_2, {\bf r}_3) 
= E \cdot \Psi({\bf r}_1, {\bf r}_2, {\bf r}_3)$, where $E < 0$ and the 
non-relativistic Hamiltonian of the three-body system which is written in the 
form of Eq.(\ref{Ham1}). To determine the highly accurate solutions of the 
non-relativistic Schr\"odinger equation with $E < 0$ in this study we apply the 
exponential variational expansion in the relative/perimetric three-body 
coordinates. The explicit form of the exponential variational expansion in 
perimetric coordinates is
\begin{eqnarray}
 \Psi_{LM} = \frac{1}{2} (1 + \kappa \hat{P}_{21}) \sum_{i=1}^{N}
 \sum_{\ell_{1}} C_{i} {\cal Y}_{LM}^{\ell_{1},\ell_{2}} ({\bf r}_{31},
 {\bf r}_{32}) \exp(-\alpha_{i} u_1 - \beta_{i} u_2 - \gamma_{i} u_3) 
 \label{equ6} 
\end{eqnarray}
where $C_{i}$ are the linear (or variational) parameters, $\alpha_i,
\beta_i, \gamma_i$ are the non-linear parameters and $L$ is the angular 
momentum of the three-body system $ab\mu$. Note that each basis function in
Eq.(\ref{equ6}) is an eigenfunction of the $L^2$ and $L_z$ operators with
eigenvalues $L(L+1)$ and $M$. This means that $\hat{L}^2 \Psi_{LM} = L (L + 
1) \Psi_{LM}$, while $M$ is the eigenvalue of the $\hat{L}_z$ operator, 
i.e. $\hat{L}_z \Psi_{LM} = M \Psi_{LM}$. The operator $\hat{P}_{21}$ in 
Eq.(\ref{equ6}) is the permutation of the two identical particles in symmetric 
three-body systems. For such systems in Eq.(\ref{equ6}) one finds $\kappa = \pm 
1$, otherwise $\kappa = 0$. In general, for the bound states of natural spatial 
parity we chose in Eq.(\ref{equ6}) $\kappa = (-1)^L$ for all symmetric muonic 
molecular ions $pp\mu, dd\mu, tt\mu$ and $\kappa = 0$ for all non-symmetric ions 
$pd\mu, pt\mu, dt\mu$.

The functions ${\cal Y}_{LM}^{\ell_{1},\ell_{2}} ({\bf r}_{31}, {\bf
r}_{32})$ in Eq.(\ref{e1}) are the bipolar harmonics \cite{Varsh} of the two
vectors ${\bf r}_{31} = r_{31} \cdot {\bf n}_{31}$ and ${\bf r}_{32} =
r_{32} \cdot {\bf n}_{32}$. The bipolar harmonics are defined as follows
\cite{Varsh}
\begin{equation}
 {\cal Y}_{LM}^{\ell_{1},\ell_{2}} ({\bf x}, {\bf y}) = x^{\ell_{1}}
 y^{\ell_{2}} \sum_{\ell_{1}, \ell_{2}} C^{LM}_{\ell_{1} m_{1};\ell_{2}
 m_{2}} Y_{\ell_{1} m_{1}} ({\bf n}_{x}) Y_{\ell_{2} m_{2}} ({\bf n}_{y})
 \label{e7}
\end{equation}
where $C^{LM}_{\ell_{1} m_{1};\ell_{2} m_{2}}$ are the Clebsch-Gordan
coefficients (see, e.g., \cite{Varsh} and \cite{Rose}) and the vectors ${\bf 
n}_{x} = \frac{{\bf x}}{x}$ and ${\bf n}_{y} = \frac{{\bf y}}{y}$ are the
corresponding unit vectors constructed for arbitrary non-zero vectors ${\bf
x}$ and ${\bf y}$. As follows from Eq.(\ref{e7}) each bipolar harmonic is
the $M-$component of the irreducible tensor of rank $L$. In actual 
calculations it is possible to use only those bipolar harmonics for which 
$\ell_{1} + \ell_{2} = L$. Note that the basis set, Eq.(\ref{equ6}), is a 
partial case of the more general exponential variational expansion in the 
relative/perimetric coordinates \cite{FrWa2011}. In particular, our 
Eq.(\ref{e1}) does not include exponents with the imaginary (or complex) 
non-linear parameters and some other factors which are needed to accelerate 
the overall convergence rate for some three-body systems, e.g., for the 
H$^{+}_2$, D$^{+}_2$ ions, helium-muonic atoms and for other `special' 
systems (for more details, see, \cite{FrWa2011} and \cite{Fro2002}). We also 
do not want to discuss here the bound states of unnatural spatial parity, 
when one needs to use in Eq.(\ref{equ6}) the bipolar harmonics for which 
$\ell_1 + \ell_2 = L + 1$.
 
\section{Spectra of bound states in muonic molecular ions}

By analyzing the bound state spectra in the six muonic molecular ions 
$pp\mu, pd\mu, pt\mu, dd\mu, dt\mu$ and $tt\mu$ one finds that they can be 
separated into three different groups \cite{Fro2001} on qualitative grounds. 
The first group includes three light muonic molecular ions $pp\mu, pd\mu$ and 
$pt\mu$. Each of these systems has two bound states: one $S(L = 0)-$state and 
one $P(L = 1)-$state. Neither of these two states is weakly bound. Note that 
each of these light muonic molecular ions contains at least one protium 
nucleus. The second group includes the two `intermediate' muonic molecular 
ions $dd\mu$ and $dt\mu$ each of which has five bound states: two $S(L = 
0)-$states, two $P(L = 1)-$states and one $D(L = 2)-$state. One of these five 
states (the excited $P^{*}(L = 1)-$state in each of these two ions) is weakly 
bound. By the formal definition the weakly bound state in a few-body system is 
a state with very small binding energy $\varepsilon$, or, in other words, with
a very small ratio of the binding and total energies, i.e., $\tau = 
\frac{\varepsilon}{E} \ll 1$, where $\tau$ is the dimensionless parameter. The 
third group contains only the heaviest muonic molecular ion $tt\mu$ which has 
six bound states (and no weakly bound states): two $S(L = 0)-$states, two $P(L 
= 1)-$states, one $D(L = 2)-$state and one $F(L = 3)-$state. 

The classification of bound state spectra in muonic molecular ions is based 
on the general theory developed in \cite{FrBi}, \cite{Fro1991} for three-body 
Coulomb systems with unit charges. This theory is based on the fact that the 
total number of bound states in any muonic molecular ion $a^{+} b^{+} \mu^{-}$ 
is determined by the lightest nucleus of the hydrogen isotope in this ion. This 
explains why only three groups of different bound state spectra can be found in 
the six muonic molecular ions mentioned above: the $p-$group, the $d-$group 
and the $t-$group. Furthermore, it must be a similarity between the spectra of 
bound states in each group: e.g., between the bound state spectra of the `protium' 
muonic molecular ions $pp\mu, pd\mu$ and $pt\mu$. Analogous similarity can be 
found for the bound state spectra of the $dd\mu$ and $dt\mu$ ions in the 
`deuterium' group. It can be shown that in such `families' of muonic molecular 
ions the symmetric ion always has the maximal binding energy \cite{Fro1991}. By 
using these similarities between the bound state spectra in each of these 
`families', one also finds a number of useful relations for the total and binding 
energies as well as for other bound state properties of different muonic molecular 
ions (see examples in \cite{FrWa2011}).

As we have mentioned above there are 22 bound states in the six muonic molecular 
ions $pp\mu, pd\mu, pt\mu, dd\mu, dt\mu$ and $tt\mu$. Nine of these states are 
the $S(L = 0)-$states, while nine others are the $P(L = 1)-$states. There are 
also three bound $D(L = 2)-$states and one bound $F(L = 3)-$state. The $F-$state 
is stable only in the heavy $tt\mu$ ion. In this study we determine the total 
energies of twenty one such bound states. At this moment we cannot perform the 
highly accurate computations of the $F(L = 3)-$state in the $tt\mu$ ion, since 
our unique code for such calculations was lost a few months ago (due to some 
problems at our local computer).     

\section{Results and discussions}

In this study all numerical computations of the $S-$ and $P-$bound states in 
muonic molecular ions in this study are performed with the use of 64 - 108 decimal 
digits per computer word \cite{Bail1}, \cite{Bail2}, allowing the total energies to 
be determined to the accuracy $\approx 1 \cdot 10^{-21} - 1 \cdot 10^{-23}$ $m.a.u.$ 
In all calculations we have assumed that all particle masses and corresponding 
conversion factors (e.g., the factor $Ry$ mentioned below) are exact. In fact, such 
assumptions are always made in the papers on highly accurate computations in few-body 
systems (see, e.g., \cite{Hili} and \cite{BaFr}). The known experimental 
uncertainties in particle masses and conversion factors are taken into account at 
the last step of calculations, when the most accurate computations are simply 
repeated for a few times with the use of different particle masses and conversion 
factors. Analogously, the expectation values of other operators are determined in 
calculations with our non-relativistic wave functions. To avoid a substantial loss 
of numerical accuracy during computations of the expectation values of some 
operators these non-relativistic wave functions must be extremely accurate.

Table I contains the total variational energies obtained for the ground and first 
`vibrationally' excited $S(L = 0)-$states of the symmetric and non-symmetric 
muonic molecular ions $pp\mu, pd\mu, pt\mu, dd\mu, dt\mu$ and $tt\mu$. Table II 
includes the total energies for the rotationally and vibrarionally excited $P(L = 
1)-$ and $P^{\ast}(L = 1)-$states of these six muonic molecular ions. In these 
two Tables and everywhere below the upper index `$\ast$' is used to designate the 
vibrationally excited state with the same angular momentum $L$. Table III contains 
the total energies of the $D(L = 2)-$states in the three heavy muonic molecular 
ion $dd\mu, tt\mu$ and $dt\mu$. The results from Table III have been obtained with 
the use of the standard Fortran with the quadruple precision accuracy (30 decimal 
digits per computer word). Also, in calculations of the bound $D(L = 2)-$states we 
did not applied our two-stage optimization of the non-linear parameters in the 
trial wave functions. Therefore, the energies from Table III  are less accurate 
than analogous energies from Tables I and II. 

In Tables I - III the notation $\infty$ stands for the total energy which 
corresponds to the infinite number of basis function, i.e. $N = \infty$ in 
Eq.(\ref{e1}). The asymptotic formula for the total energy takes the following 
four-parameteric form
\begin{equation}
 E(N_i) = E(\infty) + \frac{A_1}{N^{\gamma}_{i}} +
 \frac{A_2}{N^{\gamma+1}_{i}} \label{asympt}
\end{equation}
where $E(\infty), A_1, A_2$ and $\gamma$ are the four parameters which are 
determined by using the results of highly accurate calculations of the total 
energies with the different numbers of basis functions $N_i$ (see Eq.(\ref{equ6})). 
To determine four parameters in Eq.(\ref{asympt}) one needs to know at least four 
total energies $E(N_i)$ obtained from the direct numerical calculations. In general, 
the asymptotic value of the total energy contains one/two correct decimal digit(s) 
extra. Moreover, the exact coincidence of some decimal digits in the $E(N_i)$ and 
$E(\infty)$ energies in Eq.(\ref{asympt}) allow us to confirm the total number of 
stable decimal digits in the final energies.  

Highly accurate calculations of the total energies and other bound state properties 
for all known bound states in muonic molecular ions is an important and actual 
scientific problem. As mentioned above in this study we use the improved values of 
particle masses known from recent high energy experiments. Our results presented in 
Tables I - III provide answers for a number of actual questions. For instance, as is 
well known the bound $P^{\ast}(L = 1)-$states in the $dd\mu$ and $dt\mu$ ions are 
very weakly bound. By using the corresponding energies from Table II and improved 
masses of the muon, deuterium and tritium nuclei (see above) one finds the `improved' 
binding energies of the $P^{\ast}(L = 1)-$states in the $dd\mu$ and $dt\mu$ ions:
\begin{equation}
 \varepsilon(dd\mu; P^{\ast}(L = 1)) = -1.9749828376301(2) eV
\end{equation}
and 
\begin{equation}
 \varepsilon(dt\mu; P^{\ast}(L = 1)) = -0.6603325645(2) eV
\end{equation} 
where we have used the following conversion factor $Ry = 27.211385060 \cdot
\Bigl( \frac{m_{\mu}}{m_e} \Bigl)$ from muon-atomic units to electron volts
(1 $eV$ = 1.602176487(40) $\times 10^{-19}$ $J$). Note that such a recalculation 
from muon-atomic units to electron volts requires the knowledge of the electron 
mass $m_e$ ($m_e$ = 0.510 998 910 $MeV/c^2$) and Rydberg constant $Ry$ (or conversion
factor), The binding energies of any other bound state in muonic molecular ions can 
be evaluated analogously. These values can be compared with the binding energies
$\varepsilon(dd\mu; P^{\ast}(L = 1))$ and $\varepsilon(dt\mu; P^{\ast}(L = 1))$
determined in \cite{FrWa2011}.

Another interesting problem is to study the changes in the bound state properties
of these muonic molecular ions which are directly related to the mass variations. 
The results of these calculations (in muon-atomic units) can be found in Table IV 
for some of
the properties. The results from this Table can be compared with analogous results
from Table 8 given in \cite{FrWa2011}. Such a comparison shows the effect of mass
variation for the bound states properties which are different from the total and 
binding energies. By working with Table IV we have found the numerical mistake in
the $\langle r^2_{21} \rangle$ expectation value for the $pd\mu$ ion (extra `1' 
was added from the left side to this result in Table 8 in \cite{FrWa2011}). The 
mass corrections to the total energies can be obtained by subtracting our results 
from Table I - III from the corresponding results given in Tables 1 - 5 in 
\cite{FrWa2011}. Since the total energies in all these Tables are given in 
muon-atomic units, then we do not need to use any additional conversion factor. 

\section{Hyperfine structure of the ground states in the $pd\mu, pt\mu$ and $dt\mu$ 
         ions}

In this Section we analyze the hyperfine structure and determine the hyperfine 
structure splitting of the bound $S(L = 0)$-states in the non-symmetric 
muonic molecular ions $pd\mu, pt\mu$ and $dt\mu$. As is mentioned above there 
are four bound $S(L = 0)-$states in these three ions: three ground $S(L = 
0)-$states (one in each of these ions) and one excited $S(L = 0)-$state in the 
heavy $dt\mu$ ion. In this Section we want to investigate the hyperfine structure 
and determine the hyperfine structure splitting for each of these bound states 
by using highly accurate expectation values of the delta-functions obtained in our
highly accurate numerical computations (see above). 

The general formula for the hyperfine structure splitting $(\Delta H)_{h.s.}$ (or 
hyperfine splitting, for short) in the case of an arbitrary three-body system is 
written as the sum of the three following terms. Each of these terms is 
proportional to the product of the factor $\frac{2 \pi}{3} \alpha^2$ and 
expectation value of the corresponding (interparticle) delta-funtion. The third 
(additional) factor contains the corresponding $g-$factors (or hyromagnetic 
ratios) and scalar product of the two spin vectors. For instance, for the $pd\mu$ 
ion this formula takes the form (in atomic units) (see, e.g., \cite{LLQ}, 
\cite{Fro02})
\begin{eqnarray}
 (\Delta H)_{h.s.} = \frac{2 \pi}{3} \alpha^2 \frac{g_p g_d}{m^2_p}
 \langle \delta({\bf r}_{pd}) \rangle ({\bf s}_p \cdot {\bf s}_d)+
 \frac{2 \pi}{3} \alpha^2 \frac{g_p g_{\mu}}{m_p m_{\mu}}
  \langle \delta({\bf r}_{p\mu}) \rangle ({\bf s}_p \cdot {\bf s}_{\mu}) 
 \nonumber \\
 + \frac{2 \pi}{3} \alpha^2 \frac{g_d g_{\mu}}{m_p m_{\mu}}
  \langle \delta({\bf r}_{d\mu}) \rangle ({\bf s}_d \cdot {\bf s}_{\mu})
 \label{es1}
\end{eqnarray}
where $\alpha = \frac{e^2}{\hbar c}$ is the fine structure constant, $m_{\mu}$ 
and $m_p$ are the muon and proton masses, respectively. The factors $g_{\mu}, 
g_{p}$ and $g_{d}$ are the corresponding $g-$factors. The expression for
$(\Delta H)_{h.s.}$ is, in fact, an operator in the total spin space which has 
the dimension $(2 s_p + 1) (2 s_d + 1) (2 s_{\mu} + 1) = 12$. In our 
calculations we have used the following numerical values for the constants and 
factors from Eq.(\ref{es1}): $\alpha = 7.297352586 \cdot 10^{-3}, m_p = 
1836.152701 m_e, m_{\mu} = 206.768262 m_e$ and $g_{\mu} = -2.0023218396$. The 
$g-$factors for the proton and deuteron are deteremined from the formulas: $g_p 
= \frac{{\cal M}_p}{I_p}$ and $g_d = \frac{{\cal M}_d}{I_d}$, where ${\cal M}_p 
= 2.792847386$ and ${\cal M}_d = 0.857438230$ are the magnetic moments (in 
nuclear magnetons) of the proton and deuteron, respectively. The spin of the 
proton and deuteron is designated in Eq.(\ref{es1}) as $I_p = \frac12$ and $I_d 
= 1$.  

The analogous formula for the hyperfine structure splitting in the $pt\mu$ ion 
takes the form 
\begin{eqnarray}
 (\Delta H)_{h.s.} = \frac{2 \pi}{3} \alpha^2 \frac{g_p g_t}{m^2_p}
 \langle \delta({\bf r}_{pt}) \rangle ({\bf s}_p \cdot {\bf s}_t) +
 \frac{2 \pi}{3} \alpha^2 \frac{g_p g_{\mu}}{m_p m_{\mu}}
  \langle \delta({\bf r}_{p\mu}) \rangle ({\bf s}_p \cdot {\bf s}_{\mu}) 
 \nonumber \\
 + \frac{2 \pi}{3} \alpha^2 \frac{g_t g_{\mu}}{m_p m_{\mu}}
  \langle \delta({\bf r}_{t\mu}) \rangle ({\bf s}_t \cdot {\bf s}_{\mu})
 \label{es2}
\end{eqnarray}
where $g_t = \frac{{\cal M}_t}{I_t}$, where ${\cal M}_t = 2.9789624775$ is the
magnetic moment of the triton expressed in the nuclear magnetons and $I_t = 
\frac12$ is the spin of the triton (or tritium nucleus). The formula for the 
hyperfine structure splitting in the $dt\mu$ ion is
\begin{eqnarray}
 (\Delta H)_{h.s.} = \frac{2 \pi}{3} \alpha^2 \frac{g_d g_t}{m^2_p}
 \langle \delta({\bf r}_{dt}) \rangle ({\bf s}_d \cdot {\bf s}_t) +
 \frac{2 \pi}{3} \alpha^2 \frac{g_d g_{\mu}}{m_p m_{\mu}}
  \langle \delta({\bf r}_{d\mu}) \rangle ({\bf s}_d \cdot {\bf s}_{\mu}) 
 \nonumber \\
 + \frac{2 \pi}{3} \alpha^2 \frac{g_t g_{\mu}}{m_p m_{\mu}}
  \langle \delta({\bf r}_{t\mu}) \rangle ({\bf s}_t \cdot {\bf s}_{\mu})
 \label{es3}
\end{eqnarray}
where all values are defined above. The same formula can be applied to determine 
the hyperfine structure spllitting in the excited $S(L = 0)-$state of the $dt\mu$ 
ion. The only difference in the hyperfine structure splittings determined for the 
ground and excited states of the $dt\mu$ ion can be related with the expectation 
values of interparticle delta-functions. 

In our computations of the muonic molecular ions performed recently \cite{FrWa2011}
we have determined the expectation values of all delta-functions which are needed 
in Eqs.(\ref{es1}) - (\ref{es3}). The corresponding expectation values are shown in 
Table I. These values have been determined in muon atomic units where $m_{\mu} = 1, 
\hbar = 1, e = 1$. They must be re-calculated to the regular atomic units ($m_e = 
1, \hbar = 1, e = 1$) which are used in the formulas, Eqs.(\ref{es1}) - (\ref{es3}),
to determine the hyperfine structure splittings. In these calculations we have used 
the trial wave functions with N = 3300, 3500, 3700 and 3840 exponential basis 
functions (for more details, see \cite{FrWa2011}). The expectation values of all
interparticle delta-functions computed for the ground $S(L = 0)-$state of the 
$pd\mu$ ion are shown in Table I. The overall convergence rates of the 
delta-functions computed for each bound state in the $pt\mu$ and $dt\mu$ ions are 
very similar to the results shown in Table V.   

These expectation values of the $\delta({\bf r}_{ij})$ functions were used in the
formulas Eqs.(\ref{es1}) - (\ref{es3}) to determine the hyperfine structure splittings 
of the bound $S(L = 0)-$states of the $pd\mu, pt\mu$ and $dt\mu$ ions. Numerical 
values of the corresponding hyperfine structure splittings can be found in Tables VI
and VII. Note that these values are usually given in $MHz$, while the values of 
$(\Delta H)_{h.s.}$  which follow from Eqs.(\ref{es1}) - (\ref{es3}) are expressed in 
atomic units. To re-calculate them from atomic units to $MHz$ the conversion factor 
6.57968392061 $\cdot 10^9$ $MHz/a.u.$ was used \cite{CRC}.

In general, the $pd\mu$ and $dt\mu$ ions have similar hyperfine structure. In
particular, in each of these ions one finds twelve spin states which are separated 
in the four following groups: (1) the group with $J = 2$ (five states), (2) the 
group with $J = 1$ (three states), (3) the group of one state with $J = 0$ (one 
state) and (4) the group with $J = 1$ (three states). Here and everywhere below 
the notation $J$ stands for the total spin (or total momentum, for the $S(L = 
0)-$states) of the three-body ion. The states with $J = 2$ have the maximal energy, 
while the energy of the states from the fourth gourp is minimal. The corresponding 
splittings $\Delta_{12}, \Delta_{23}$ and $\Delta_{34}$ can be found in Table VI 
for each bound state in the $pd\mu$ and $dt\mu$ ions.

The hyperfine structure of the ground state in the $pt\mu$ ion is completely 
different (see Table VII), since the spin of the triton equals $\frac12$, while the 
spin of the deuteron (or deuterium nucleus) equals 1. In the case of the ground state 
in the $pt\mu$ ion one finds only eight spin states which are separated into three 
different groups: (1) the group of four states with $J = \frac32$, (2) the group of 
two states with $J = \frac12$ and (3) the group of two states with $J = \frac12$. 
The group (1) has the maximal energy, while the energy of the states from the third 
group is minimal. The corresponding values of the hyperfine structure splittings in
the ground state of the $pt\mu$ ion are $\Delta_{12}$ = 1.3400149$\cdot 10^7$ $MHz$ 
and $\Delta_{23}$ = 3.3518984$\cdot 10^7$ $MHz$.

In this Section we have investigated the hyperfine structure and determine the hyperfine
structure splitting in the bound $S(L = 0)-$states of the $pd\mu, pt\mu$ and $dt\mu$ 
ions. The first excited $S(L = 0)-$state in the $dt\mu$ ion is traditionally designated 
by an additional asterisk, i.e. $(dt\mu)^*$. In such calculations we used the highly 
accurate expectation values of all interparticle delta-functions obtained in recent 
computations \cite{FrWa2011}. In general, it is very interesting to compare the numerical 
values of the hyperfine structure splittings $\Delta_{12}, \Delta_{23}$ and $\Delta_{34}$ 
for different muonic ions (see Table VI).  

\section{Hyperfine structure of the bound $S(L = 0)-$states in the symmetric muonic 
         molecular ions}

In this Section we consider the hyperfine structure splitting in the symmetric muonic 
molecular ions $pp\mu, dd\mu$ and $tt\mu$. As is well known there are five bound $S(L 
= 0)-$states in these (symmetric) muonic molecular ions. The ground states are stable 
in each of these ions, while the excited $S(L = 0)-$states are stable only in the 
heavy $dd\mu$ and $tt\mu$ ions. In general, the analysis of the hyperfine structure 
in symmetric systems is slightly more complicated than analogous analysis for 
non-symmetric systems/ions. On the other hand, the arising hyperfine structure 
is relatively simple and can be explained by using a few transparent physical 
ideas.

The general formula for the hyperfine structure splitting (or hyperfine 
splitting, for short) for an arbitrary three-body muonic molecular ion $aa\mu$ 
is written in the following form (in atomic units) (see, e.g., \cite{LLQ})
\begin{eqnarray}
 (\Delta H)_{h.s.} = \frac{2 \pi}{3} \alpha^2 \frac{g_a g_a}{m^2_p}
 \langle \delta({\bf r}_{aa}) \rangle ({\bf s}_a \cdot {\bf s}_a)+
 \frac{2 \pi}{3} \alpha^2 \frac{g_a g_{\mu}}{m_p m_{\mu}}
  \langle \delta({\bf r}_{a\mu}) \rangle ({\bf s}_a \cdot {\bf s}_{\mu}) 
 \nonumber \\
 + \frac{2 \pi}{3} \alpha^2 \frac{g_a g_{\mu}}{m_p m_{\mu}}
  \langle \delta({\bf r}_{a\mu}) \rangle ({\bf s}_a \cdot {\bf s}_{\mu})
 \label{e1}
\end{eqnarray}
where $\alpha = \frac{e^2}{\hbar c}$ is the fine structure constant, $m_{\mu}$ 
and $m_p$ are the muon and proton masses, respectively. The factors $g_{\mu}$
and $g_{a}$ are the corresponding $g-$factors. The expression for
$(\Delta H)_{h.s.}$ is, in fact, an operator in the total spin space which has 
the dimension $(2 s_a + 1)^2 (2 s_{\mu} + 1)$. Since the second and third terms 
in Eq.(1) are identical, then we can reduce Eq.(\ref{e1}) to the 
form
\begin{eqnarray}
 (\Delta H)_{h.s.} = \frac{2 \pi}{3} \alpha^2 \frac{g_a g_a}{m^2_p}
 \langle \delta({\bf r}_{aa}) \rangle ({\bf s}_a \cdot {\bf s}_a)+
 \frac{2 \pi}{3} \alpha^2 \frac{g_a g_{\mu}}{m_p m_{\mu}}
 \langle \delta({\bf r}_{a\mu}) \rangle ({\bf S}_{aa} \cdot 
 {\bf s}_{\mu}) \label{e2}
\end{eqnarray}
where ${\bf S}_{aa} = ({\bf s}_a + {\bf s}_a)$ is the total spin of the pair
of identical particles (the two nuclei of the hydrogen isotopes), i.e. $p, d$ and 
$t$.

The formula, Eq.(\ref{e2}), allows one to make a few qualitative predictions about 
the hyperfine structure of the symmetric muonic molecular ions. First, it is clear
that the classifications of the levels of hyperfine structure must be based on
the total spin of the two `symmetric' nuclei ${\bf S}_{aa}$. The absolute values of 
the spin ${\bf S}_{aa}$ are always non-negative integer numbers, i.e. $\mid {\bf 
S}_{aa} \mid = 0, 1, 2, \ldots$. For instance, in the case of two protons $p$ and/or 
two tritons $t$ one finds $\mid {\bf S}_{aa} \mid = 0, 1$, while for the two 
deuterons $\mid {\bf S}_{aa} \mid = 0, 1, 2$. The hyperfine energy of this state 
with $J = 0$ is determined only by the first term in Eq.(\ref{e2}) which is very 
small, since the expectation values $\langle \delta({\bf r}_{aa}) \rangle$ in all 
muonic molecular ions are very small. As follows from actual computations all these
values are less than $4 \cdot 10^{-5}$ (in muon atomic units). Briefly, we can say 
that the energy of this hyperfine state (with $J = \frac12$) is determined by the 
spin-spin interaction between the two heavy nuclei (muon's spin does not contribute). 
The overall contribution from the first term in Eq.(\ref{e2}) rapidly (exponentially) 
decreases when the mass of the heavy particle increases. Formally, the first term in 
Eq.(\ref{e2}) is very small already for the $pp\mu$ ion. However, for the $dd\mu$ 
and $tt\mu$ ions its contibution is negligible. This means that in the first 
approximation the hyperfine structure of the symmetric muonic molecular ions can be 
explained by using only one term for the muon-nuclear spin interaction. This leads 
to some `additional' symmetry observed for the actual levels of hyperfine structure 
of heavy ions (see below).   

As is well known the spin of the negatively charged muon $\mu^{-}$ equals $\frac12$ 
and the spins of the proton $p$ and triton $t$ also equal $\frac12$. Therefore, the 
hyperfine structure of the $pp\mu$ and $tt\mu$ ions must include eight levels which 
form three following groups: (1) the group of four spin states with $J = \frac32$, 
(2) the upper group of two states with $J = \frac12$ and (3) the lower group of two 
states with $J = \frac12$. The classification is true for the excited $S(L = 0)-$state 
in the $tt\mu$ ion. Here and everywhere below the notation $J$ stands for the total 
spin (or total momentum) of the three-body system/ion, since ${\bf J} = {\bf L} + {\bf 
S} = {\bf S}$ for the $S(L = 0)-$states).

The hyperfine structure of the $dd\mu$ ion is substantially different. In the $dd\mu$ 
ion one finds eighteen levels of hyperfine structure which are separated into five 
different groups: one group with $J = \frac52$ (six states), two different groups of 
states (upper and lower groups) with $J = \frac32$ (four states in each), two 
different groups of states (upper and lower groups) with $J = \frac12$ (two states 
in each). 

In our calculations of the hyperfine structure we have used the following numerical 
values for the constants and factors in Eq.(\ref{e2}): $\alpha = 7.297352586 
\cdot 10^{-3}, g_{\mu} = -2.0023218396$ \cite{CRC} and $m_p = 1836.152701 m_e, m_{\mu} 
= 206.768262 m_e$. The $g-$factors for the proton, deuteron and triton are deteremined 
from the formulas: $g_p = \frac{{\cal M}_d}{I_p}, g_d = \frac{{\cal M}_d}{I_d}$ and 
$g_t = \frac{{\cal M}_t}{I_t}$, where ${\cal M}_p = 2.792847386, {\cal M}_d = 
0.857438230$ and ${\cal M}_t = 2.97896247745$ are the magnetic moments (in nuclear 
magnetons) of the proton, deuteron and triton, respectively. Here the spins of the 
proton, deuteron and triton are designated in by the letter $I$ with the corresponding 
index: $I_p = \frac12, I_d = 1$ and $I_t = \frac12$. In Eqs.(\ref{e1}) - (\ref{e2}) 
these values are designated differently. In highly accurate computations of the 
expectation values of delta-functions we have used the following masses of the 
deuteron and triton: $m_d$ = 3680.483014 $m_e$ and $m_t$ = 5496.92158 $m_e$. These
masses are often used in modern highly accurate calculations of muonic molecular ions
(see, e.g., \cite{FrWa2011}). 

The convergence of the expectation values of the nuclear-nuclear (or $pp-$) and 
nuclear-muonic (or $p\mu-$) delta-functions is illustrated in Table VIII for the $pp\mu$ 
ion. The convergence of these expectation values for other bound $S(L = 0)-$states in 
the $dd\mu$ and $tt\mu$ ions is very similar to the results presented in Table VIII for 
the $pp\mu$ ion. The hyperfine structure and energy splittings between the corresponding 
levels for all five bound $S(L = 0)-$states in the three muonic molecular ions $pp\mu, 
dd\mu$ and $tt\mu$ can be found in Tables IX and X. In atomic physics these values are 
traditionally given in $MHz$. The corresponding conversion factor is 6.57968392061$\cdot 
10^9$ $MHz/a.u.$ In Tables IX and X the excited states are designated by the asterisk 
used as the upper index, e.g., $(dd\mu)^{*}$ and $(tt\mu)^{*}$. Such a system of notation 
is often used for muonic molecular ions. 

Tables IX and X contain both the energies of the levels of hyperfine structure 
($\epsilon_{J}$) and hyperfine structure splitting ($\Delta(J_1 \rightarrow J_2)$). As 
we have predicted (see above) one of the hyperfine levels has a very small energy. As 
follows from Tables IX and X this level corresponds to $J=\frac12$. In the $tt\mu$ 
ion the hyperfine energies of this state are $\approx$ 11.0591 $MHz$ and $\approx$ 
12.4307 $MHz$ for the ground and first excited states, respectively. In the $dd\mu$
ion the energies of the analogous levels are 6.8996 $MHz$ and 4.7378 $MHz$, 
respectively. Briefly, this means that the overall contribution of the nuclear-nuclear 
spin interaction is very small for the both $dd\mu$ and $tt\mu$ ions. This directly 
follows from the known fact (see, e.g., \cite{Fro99}) that the expectation values of 
nuclear-nuclear delta-functions are very small. For instance, for the $dd\mu$ and 
$tt\mu$ ions the expectation values of nuclear-nuclear delta-functions are 
$\langle \delta_{dd} \rangle \approx 2.43871205 \cdot 10^{-6}$ ($m.a.u.$), 
$\langle \delta_{dd} \rangle \approx 1.67460229 \cdot 10^{-6}$ ($m.a.u.$),
$\langle \delta_{tt} \rangle \approx 2.15893994 \cdot 10^{-7}$ ($m.a.u.$) and
$\langle \delta_{tt} \rangle \approx 2.42670033 \cdot 10^{-7}$ ($m.a.u.$), for the ground 
and excited states, respectively. Finally, the observed hyperfine structure of these two 
ions is mainly (99.9999 \%) related to the muon-nuclear spin interactions only. In the 
$pp\mu$ ion the situation is slightly different, but even for this ion the overall 
contribution of the muon-nuclear spin interaction(s) is substantially larger than the 
contribution from the nuclear-nuclear spin interaction.  

\section{Conclusion}  

Thus, we have determined the total energies of the twenty one bound states in 
the six muonic molecular ions $pp\mu, pd\mu, pt\mu, dd\mu, dt\mu$ and $tt\mu$. 
The angular momentum of these bound states equals $L$ = 0, $L$ = 1 and $L$ = 2.
Our calculations of the bound state energies in this study have been performed 
with the improved particle masses known from recent high-energy experiments. The 
highly accurate wave functions of muonic molecular ions are needed to determine the 
expectation values of some operators. These expectation values are of interest in 
numerous applications related to the muonic molecular ions. Currently, all muonic 
molecular ions can  be created in real experiments and their various properties can 
be measured to very good accuracy. Therefore, we can compare the predicted (or 
computed) values of the bound state properties with their actual (or observed) 
values. In general, the analysis of these three-body systems is significantly more 
interesting and informative than the traditional analysis of the two-electron atoms 
and ions.

The hyperfine structure of all nine bound $S(L = 0)-$states of the $pp\mu, pd\mu, 
pt\mu, dd\mu, dt\mu$ and $tt\mu$ has also been investigated. In our calculations of 
hyperfine structure and hyperfine structure splittings for each of these states we 
used the highly accurate expectation values of the interparticle delta-functions. 
The hyperfine structure splittings of the ground states of each of these ions (see 
Tables VI, VII, IX and X) can directly be measured in modern experiements. 

\begin{center}
  {\bf Appendix}
\end{center}

In this Appendix we briefly discuss the history of the bound state computations 
of muonic molecular ions. Note that the interest to these ions was always closely 
related to the problems of muon-catalyzed fusion of the nuclear reactions. The 
first numerical computations of the bound states in three-bodymuonic molecular ions 
were performed by Belyaev et al in 1959 \cite{Bel}. By using a very simple adiabatic 
(but non-variational!) procedure they were able to find 20 bound states in six ions 
$pp\mu, pd\mu, pt\mu, dd\mu, dt\mu$ and $tt\mu$. Unfortunately, due to lack of good 
computers at that time the overall accuracy of the procedure used in \cite{Bel} was 
very low and the authors could not confirm the boundness of the excited $P^{*}(L = 
1)-$states (or (1,1)-states) in the $dd\mu$ and $dt\mu$ ions. It was concluded only 
that, if such states are bound, then they are very weakly bound. The binding energy 
of these two states was expected to be smaller than 4.5 $eV$, i.e. smaller than the 
binding energy of a typical hydrogen molecule. Immediately after publication of 
\cite{Bel} an intense stream of speculations started about a possible interference 
(or resonance) between the formation of excited $P^{*}(L = 1)-$states (or (1,1)-states) 
in the $dd\mu$ and $dt\mu$ muonic molecular ions and different transitions in 
surrounding molecules (see, e.g., \cite{Zel} and references there in). 

In the middle of 1960's Halpern \cite{Hal}, Carter \cite{Car1}, \cite{Car2} and 
Delves and Kalotas \cite{Del} published their papers with the results of
variational computations obtained for some bound states in the muonic molecular 
ions. In particular, Halpern \cite{Hal} considered the bound $P(L = 1)-$states 
in the symmetric $pp\mu, dd\mu, tt\mu$ ions. The paper of Delves and Kalotas has 
a great methodological interest, since 99 \% of all modern methods for 
highly accurate computations of Coulomb three-body systems are based on that work.
At the same time a number of experiments have been performed by Bystritskii
et al (see \cite{Exp1} and \cite{Exp2} and references therein). They worked with 
$\mu^{-}$-muons which were slowing down in liquid deuterium and deuterium-tritium 
mixture. Finally, it was found that one muon can catalyze approximately 10 - 20 
$(d,d)-$nuclear reactions in liquid deuterium (D$_2$) and 90 - 110 $(d,t)-$reactions 
in the liquid equimolar deuterium-tritium mixture (D$_2$ : T$_2$ = 1:1). Such very 
large numbers of nuclear reactions catalyzed by one muon can be explained only by the 
resonance (or very fast) formation of $dd\mu$ and $dt\mu$ muonic molecular ions. 
Correspondingly, the related processes were called `resonance' muon-catalyzed fusion 
of nuclear reactions, in contrast with the `regular' muon-catalyzed fusion observed 
in \cite{Alv}. 

Those experimental works produced a great interest to study the weakly-bound states 
in the $dd\mu$ and $dt\mu$ ions. The main goal of all following computations was to
determine the binding energies of the weakly-bound $P^{*}(L = 1)-$states in the 
$dd\mu$ and $dt\mu$ ions to the accuracy $\approx 1 \cdot 10^{-3}$
$eV$ (or approximately 10 $K$). At that time a large number of bound state 
computations for muonic molecular ions were performed with the use of the adiabatic 
(but non-variational!) method \cite{Vin}. The first variational computations of all
bound $S(L = 0)-$ and $P(L = 1)-$states in muonic molecular ions have been conducted 
in \cite{FrEf} and \cite{BaDr}. Later, we have substantially improved the accuracy of 
such computations \cite{FrEf85} and were able to calculate the bound $D(L = 2)-$states 
in these ions \cite{Fro86}. Since then we have performed a number of different 
computations of the bound $D(L = 2)-$states in the $dd\mu, dt\mu$ and $tt\mu$ ion. The
total energy of the $dt\mu$ ion \cite{Kam} is the only result obtained for these states 
outside of our group. However, in the middle of 1980's the masses of all particles 
involved in the muonic molecular ions were determined to much better accuracy. In addition
to this, it became finally clear that the main restriction of the resonance 
muon-catalized fusion is directly related to the muon stiking coefficient to the ${}^4$He 
nucleus, or with its inverse value which equals to the number of nuclear reactions 
catalyzed by one muon in the equimolar deuterium-tritium mixture. It appears that such a 
number (150 - 160) was very close to the value obtained earlier \cite{Exp1} and \cite{Exp2}. 
It was also shown that even 200 nuclear fusions per one muon in the equimolar 
deuterium-tritium (liquid) mixture is in 12 - 15 times less than it is needed for 
theoretical break-even and $\approx$ 55 - 65 times smaller than necessary for actual 
break-even (see discussion and references in \cite{MS}).

After these publications the overall interest to the resonance muon-catalyzed fusion 
rapidly went down. Nevertheless, in the middle of 1990's we have performed a series of 
highly accurate computations of the muonic molecular ions \cite{FrBi1}, \cite{FrBi2}. 
These works were originally stimulated by the development of the theory of bound
states in the Coulomb three-body systems with unit charges. Later, I have wrote the 
code to determine the energy and some other properties of the $F(L = 3)-$state in the 
$tt\mu$ ion. In our calculations performed around 2001 - 2003 we have used the advanced 
Fortran pre-translator written by D.H. Bailey \cite{Bai1}, \cite{Bai2}. Around that time 
another paper was published on highly accurate bound state computations of the three-body
muonic molecular ions \cite{Hil}. In \cite{FrWa2011} we have used very large basis sets 
and performed an accurate optimization of the non-llinear parameters of our method. The 
paper \cite{FrWa2011} contains the most accurate values of the total (and binding 
energies) of all 22 bound states in the six muonic molecular ions $pp\mu, pd\mu, pt\mu, 
dd\mu, dt\mu$ and $tt\mu$. In this study we wanted to recalculate some of these systems by 
using the new values of the particle masses.   
 
Note in conclusion that despite an obvious failure of the resonance muon-catalyzed 
fusion in the D:T mixture the muon-catalysis of the nuclear reactions is not a closed 
problem. In this area one still finds dozens of unsolved, approximately and wrongly solved 
problems. One of such problems is the current (very large) deviation between theoretical 
and experimental fusion rates in the $pt\mu$ ion. Other open problems include accurate 
evaluations of the fusion rates and muon stiking probabilities for some rotationally 
excited states, probabilities of excitations (and de-excitations) of muon-molecular ions, 
etc.

\newpage

   \begin{table}[tbp]
    \caption{The total energies ($E$) of the bound $S(L = 0)-$states in
             the symmetric muonic molecular ions in muon-atomic units
             ($m_{\mu} = 1, \hbar = 1, e = 1$). $N$ designates the number
             of basis functions used in Eq.(2).}
      \begin{center}
      \begin{tabular}{llll}
        \hline\hline
 $N$ & $E(pp\mu)$ & $E(dd\mu)$ & $E(tt\mu)$ \\
     \hline
 3300 & -0.494386 815212 835026 521839  & -0.531111 132193 187917 45 & -0.546374 225613 816728 844 \\

 3500 & -0.494386 815212 835026 522038  & -0.531111 135402 386449 61 & -0.546374 225613 816728 849 \\

 3700 & -0.494386 815212 835026 522184  & -0.531111 135402 386450 59 & -0.546374 225613 816728 855 \\

 3840 & -0.494386 815212 835026 522266  & -0.531111 135402 386451 22 & -0.546374 225613 816728 856 \\
     \hline
  $\infty$ & -0.494386 815212 835026 52250(10) & -0.531111 135402 386455(2) & -0.546374 225613 816728 90(3) \\
      \hline\hline
 $N$ & $E(pd\mu)$ & $E(pt\mu)$ & $E(dt\mu)$ \\
      \hline
 3300 & -0.512 711 792 481 703 484 & -0.519 880 085 704 058 459 & -0.538 594 971 709 480 710 \\

 3500 & -0.512 711 792 481 703 573 & -0.519 880 085 704 058 570 & -0.538 594 971 709 480 718 \\

 3700 & -0.512 711 792 481 703 647 & -0.519 880 085 704 058 670 & -0.538 594 971 709 480 724 \\ 

 3840 & -0.512 711 792 481 703 670 & -0.519 880 085 704 058 711 & -0.538 594 971 709 480 730 \\
     \hline
  $\infty$ & -0.512 711 792 481 703 85(4) & -0.519 880 085 704 058 87(4) & -0.538 594 971 709 480 79(2) \\
    \hline\hline
 $N$ & $E^{*}(dd\mu)$ & $E^{*}(dt\mu)$ & $E^{*}(tt\mu)$ \\
     \hline
 3300 & -0.47970 63771 01901 40596 & -0.488 065 354 215 765 737 & -0.49676 28898 97946 30625 \\

 3500 & -0.47970 63771 01901 40638 & -0.488 065 354 215 765 800 & -0.49676 28898 97946 30709 \\

 3700 & -0.47970 63771 01901 40667 & -0.488 065 354 215 765 843 & -0.49676 28898 97946 30786 \\

 3840 & -0.47970 63771 01901 40689 & -0.488 065 354 215 765 860 & -0.49676 28898 97946 30823 \\
     \hline
  $\infty$ & -0.47970 63771 01901 4080(3) & -0.488 065 354 215 765 98(4) & -0.49676 28898 97946 314(3) \\
      \hline\hline
  \end{tabular}
  \end{center}
  \end{table}
   \begin{table}[tbp]
    \caption{The total energies ($E$) of the bound $P(L = 1)-$states in
             the symmetric muonic molecular ions in muon-atomic units
             ($m_{\mu} = 1, \hbar = 1, e = 1$). $N$ designates the number
             of basis functions used in Eq.(2).}
      \begin{center}
      \begin{tabular}{llll}
        \hline\hline
 $N$ & $E(pp\mu)$ & $E(dd\mu)$ & $E(tt\mu)$ \\
     \hline
 3300 & -0.468 458 430 358 808 027 03 & -0.513 623 952 704 526 3269 & -0.533 263 445 209 533 2938 \\

 3500 & -0.468 458 430 358 808 031 55 & -0.513 623 952 704 526 3342 & -0.533 263 445 209 533 3070 \\

 3700 & -0.468 458 430 358 808 035 14 & -0.513 623 952 704 526 3387 & -0.533 263 445 209 533 3189 \\

 3840 & -0.468 458 430 358 808 037 51 & -0.513 623 952 704 526 3407 & -0.533 263 445 209 533 3254 \\
     \hline
  $\infty$ & -0.468 458 430 358 808 045(3) & -0.513 623 952 704 526 39(3) & -0.533 263 445 209 533 38(3) \\
      \hline\hline
 $N$ & $E(pd\mu)$ & $E(pt\mu)$ & $E(dt\mu)$ \\
      \hline
 3300 & -0.490 664 164 603 504 64 & -0.499 492 024 990 190 10 & -0.523 191 452 003 587 60 \\

 3500 & -0.490 664 164 603 507 57 & -0.499 492 024 990 191 53 & -0.523 191 452 003 588 67 \\

 3700 & -0.490 664 164 603 510 41 & -0.499 492 024 990 192 53 & -0.523 191 452 003 589 54 \\

 3840 & -0.490 664 164 603 511 76 & -0.499 492 024 990 192 96 & -0.523 191 452 003 590 17 \\
     \hline
  $\infty$ & -0.490 664 164 603 515(1) & -0.499 492 024 990 195(1) & -0.523 191 452 003 593(1) \\
    \hline\hline
 $N$ & $E^{*}(dd\mu)$ & $E^{*}(dt\mu)$ & $E^{*}(tt\mu)$ \\
     \hline
 3300 & -0.473 686 731 121 137 629 & -0.481 991 527 054 2451 & -0.489 908 663 057 013 5630 \\

 3500 & -0.473 686 731 121 137 765 & -0.481 991 527 054 3644 & -0.489 908 663 057 013 6844 \\

 3700 & -0.473 686 731 121 137 946 & -0.481 991 527 054 4505 & -0.489 908 663 057 013 8018 \\

 3840 & -0.473 686 731 121 138 063 & -0.481 991 527 054 4894 & -0.489 908 663 057 013 8571 \\
     \hline
  $\infty$ & -0.473 686 731 121 138 5(3) & -0.481 991 527 054 9(3) & -0.489 908 663 057 014 5(2) \\
      \hline\hline
  \end{tabular}
  \end{center}
  \end{table}
   \begin{table}[tbp]
    \caption{The total energies ($E$) of the bound $D(L = 2)-$states of the
             the $dd\mu, tt\mu$ and $dt\mu$ muonic molecular ions in muon
             atomic units. $N$ designates the number of basis functions
             used in Eq.(2).}
      \begin{center}
      \begin{tabular}{lllll}
        \hline\hline
 $N$ & $E(dd\mu)$ & $E(tt\mu)$ & $N$ & $E(dt\mu)$ \\
      \hline
 2800 & -0.488 708 327 4382 & -0.512 568 647 4651 & 3900 & -0.500 118 078 7334 \\

 3000 & -0.488 708 327 5083 & -0.512 568 647 6583 & 4200 & -0.500 118 078 7337 \\

 3200 & -0.488 708 327 5460 & -0.512 568 647 8022 & 4500 & -0.500 118 078 7338 \\

 3400 & -0.488 708 327 5584 & -0.512 568 647 9152 & 4800 & -0.500 118 078 7338 \\

 3600 & -0.488 708 327 5715 & -0.512 568 648 0085 & 5100 & -0.500 118 078 7338 \\
     \hline
  $\infty$ & -0.488 708 327 598(3) & -0.512 568 648 210(5) & $\infty$ &
        -0.500 118 078 7338(1) \\
     \hline\hline
  \end{tabular}
  \end{center}
  \end{table}
\newpage
  \begin{table}[tbp]
   \caption{The bound state properties $X$ computed for the ground
            $S(L = 0)-$state and excited $S^{*}(L = 0)-$state in the
            $pd\mu$ and $dt\mu$ muonic molecular ions (in muon-atomic
            units).}
     \begin{center}
     \begin{tabular}{cccc}
        \hline\hline
$X$ & $pd\mu (S(L = 0)-$state) & $dt\mu (S(L = 0)-$state) & $dt\mu (S^{*}(L =
       0)-$state) \\
        \hline
$\langle r_{31}^{-1} \rangle$ &
0.6411463600638491(1) &
0.7227000026976390(3) &
0.5146887255965181(3) \\

$\langle r_{32}^{-1} \rangle$ &
0.7533736114443716(1) &
0.7583156054974041(3) &
0.7053753065541911(3) \\

$\langle r_{21}^{-1} \rangle$ & 
0.3690963865448171(3) &
0.4038256647760819(5) & 
0.2439333237191783(6) \\
       \hline
$\langle r_{31} \rangle$ &
2.451487643610344(2) &
2.117912271227347(3) &
3.933236044506724(3) \\

$\langle r_{32} \rangle$ &
2.087699160470998(2) &
2.023720516217468(3) &
2.738751054881910(3) \\

$\langle r_{21} \rangle$ &
3.100710462458351(3) &
2.747914171742117(5) &
5.161229304527515(5) \\
        \hline
$\langle r^2_{31} \rangle$ &
8.033494559453071(3) &
5.881854047519130(4) &
22.39719714570313(5)\\

$\langle r^2_{21} \rangle$ &
10.829021567566262(4) &
8.2873255690653121(6) &
30.631304593819174(6) \\

$\langle r^3_{32} \rangle$ &
20.6547098820345(2) &
17.4696971262928(3) &
65.2551287149941(3) \\

$\langle r^3_{21} \rangle$ &
42.147966883621(4) &
27.208344987497(5) &
201.45182645373(6) \\
         \hline
$\langle -\frac12 \nabla^2_1 \rangle$ &
0.2806191821887450(7) &
0.3910764626134503(9) &
0.3836327048408983(9) \\

$\langle -\frac12 \nabla^2_2 \rangle$ &
0.3674608971218589(7) &
0.4218837688718115(9) &
0.4476331876944995(9) \\

$\langle -\frac12 \nabla^2_3 \rangle$ &
0.46041133009768349(4) &
0.50069529322880649(5) &
0.50035758500774130(6) \\
      \hline\hline
  \end{tabular}
  \end{center}
  \end{table}
%
%

\begin{table}[tbp]
    \caption{The convergence of the $\langle \delta_{32} \rangle, 
             \langle \delta_{31} \rangle$ and $\langle \delta_{31} \rangle$
             expectation values for the ground (bound) $S(L = 0)-$state of 
             the $pd\mu$ molecular ion (in muon-atomic units).}
      \begin{center}
      \begin{tabular}{llll}
        \hline\hline
 $N$ & $\langle \delta_{32} \rangle$ & $\langle \delta_{31} \rangle$ & $\langle \delta_{21} 
 \rangle$ \\
      \hline
  3300 & 1.73456203087$\cdot 10^{-1}$ & 1.17709732798$\cdot 10^{-1}$ & 1.46169407$\cdot 10^{-5}$ \\

  3500 & 1.73456202965$\cdot 10^{-1}$ & 1.17709733128$\cdot 10^{-1}$ & 1.46169370$\cdot 10^{-5}$ \\

  3700 & 1.73456202754$\cdot 10^{-1}$ & 1.17709733014$\cdot 10^{-1}$ & 1.46169377$\cdot 10^{-5}$ \\

  3840 & 1.73456202768$\cdot 10^{-1}$ & 1.17709733160$\cdot 10^{-1}$ & 1.46169383$\cdot 10^{-5}$ \\
       \hline\hline
   \end{tabular}
   \end{center}
   \end{table}
\begin{table}[tbp]
    \caption{The levels of hyperfine structure $\epsilon$ and hyperfine structure splittings 
             $\Delta$ in the bound $S(L = 0)-$states of the $pd\mu$ and $dt\mu$ ions 
             (in $MHz$).}
      \begin{center}
      \begin{tabular}{lllll}
        \hline\hline
$\epsilon_{J=2}(pd\mu)$ &  1.2519350851$\cdot 10^7$ & ----- & ------------ \\ 

$\epsilon_{J=1}(pd\mu)$ &  9.3058194294$\cdot 10^6$ & $\Delta(J=2 \rightarrow J=1)$ & 3.2135314217$\cdot 10^6$ \\ 

$\epsilon_{J=0}(pd\mu)$ & -2.1222395094$\cdot 10^7$ & $\Delta(J=1 \rightarrow J=0)$ & 3.0528214524$\cdot 10^7$ \\

$\epsilon_{J=1}(pd\mu)$ & -2.3097272483$\cdot 10^7$ & $\Delta(J=0 \rightarrow J=1)$ & 1.8748773889$\cdot 10^6$ \\
      \hline
      \hline
$\epsilon_{J=2}(dt\mu)$ &  1.8919590437$\cdot 10^7$ & ----- & ------------ \\ 

$\epsilon_{J=1}(dt\mu)$ &  1.5985479092$\cdot 10^6$ & $\Delta(J=2 \rightarrow J=1)$ & 2.9341113453$\cdot 10^6$ \\ 

$\epsilon_{J=0}(dt\mu)$ & -3.4439378258$\cdot 10^7$ & $\Delta(J=1 \rightarrow J=0)$ & 5.0424857350$\cdot 10^7$ \\

$\epsilon_{J=1}(dt\mu)$ & -3.6038337067$\cdot 10^7$ & $\Delta(J=0 \rightarrow J=1)$ & 1.5989588085$\cdot 10^6$ \\
      \hline
      \hline
$\epsilon_{J=2}(dt\mu)^{*}$ &  1.8609555434$\cdot 10^7$ & ----- & ------------ \\ 

$\epsilon_{J=1}(dt\mu)^{*}$ &  1.6554331952$\cdot 10^6$ & $\Delta(J=2 \rightarrow J=1)$ & 2.0552234821$\cdot 10^6$ \\ 

$\epsilon_{J=0}(dt\mu)^{*}$ & -3.4859818025$\cdot 10^7$ & $\Delta(J=1 \rightarrow J=0)$ & 5.1414149977$\cdot 10^7$ \\

$\epsilon_{J=1}(dt\mu)^{*}$ & -3.5950318334$\cdot 10^7$ & $\Delta(J=0 \rightarrow J=1)$ & 1.0905003094$\cdot 10^6$ \\
      \hline\hline
   \end{tabular}
   \end{center}
   \end{table}
\begin{table}[tbp]
    \caption{The levels of hyperfine structure $\epsilon$ and hyperfine structure splittings
             $\Delta$ in the ground $S(L = 0)-$state of the $pt\mu$ ion (in $MHz$).}
      \begin{center}
      \begin{tabular}{lllll}
        \hline\hline
$\epsilon_{J=\frac32}(pt\mu)$ &  1.5079820356$\cdot 10^7$ & ----- & ------------ \\ 

$\epsilon_{J=\frac12}(pt\mu)$ &  1.6796717260$\cdot 10^6$ & $\Delta(J=\frac32 \rightarrow J=\frac12)$ & 1.3400148630$\cdot 10^7$ \\ 

$\epsilon_{J=\frac12}(pt\mu)$ & -3.1839312439$\cdot 10^7$ & $\Delta(J=\frac12 \rightarrow J=\frac12)$ & 3.3518984165$\cdot 10^7$ \\
      \hline
      \hline
   \end{tabular}
   \end{center}
   \end{table}


\begin{table}[tbp]
    \caption{The convergence of the $\langle \delta_{p\mu} \rangle$ and $\langle 
             \delta_{pp} \rangle$ expectation values for the ground (bound) $S(L 
             = 0)-$state of the $pp\mu$ molecular ion (in muon-atomic units).}
      \begin{center}
      \begin{tabular}{lll}
        \hline\hline
 $N$ & $\langle \delta_{31} \rangle$ & $\langle \delta_{21} \rangle$ \\  
      \hline
  3300 & 1.315008614364$\cdot 10^{-1}$ & 3.9370034861$\cdot 10^{-5}$ \\

  3500 & 1.315008614369$\cdot 10^{-1}$ & 3.9370034722$\cdot 10^{-5}$ \\

  3700 & 1.315008614374$\cdot 10^{-1}$ & 3.9370034782$\cdot 10^{-5}$ \\

  3840 & 1.315008614378$\cdot 10^{-1}$ & 3.9370034773$\cdot 10^{-5}$ \\
       \hline\hline
   \end{tabular}
   \end{center}
   \end{table}
\begin{table}[tbp]
    \caption{The hyperfine structure and hyperfine structure splitting of the 
             bound $S(L = 0)-$states of the $pp\mu$ and $tt\mu$ ions (in $MHz$).}
      \begin{center}
      \begin{tabular}{lllll}
        \hline\hline
      & $pp\mu$ & $tt\mu$ & $(tt\mu)^{*}$\\
      \hline
$\epsilon_{J=\frac32}$ &  1.256448515$\cdot 10^7$ &  1.736310113$\cdot 10^7$ &  1.510018118$\cdot 10^7$ \\

$\epsilon_{J=\frac12}$ &  1.772596177$\cdot 10^3$ &  1.105911127$\cdot 10^1$ &  1.243070661$\cdot 10^1$ \\

$\epsilon_{J=\frac12}$ & -2.513074289$\cdot 10^7$ & -3.472621366$\cdot 10^7$ & -3.020037480$\cdot 10^7$ \\
      \hline
$\Delta(\frac32 \rightarrow \frac12)$ & 1.256271251$\cdot 10^7$ & 1.735309024$\cdot 10^7$ & 1.510016875$\cdot 10^7$ \\

$\Delta(\frac12 \rightarrow \frac12)$ & 2.513251549$\cdot 10^7$ & 3.472622472$\cdot 10^7$ & 3.020038723$\cdot 10^7$ \\
       \hline\hline
   \end{tabular}
   \end{center}
   \end{table}
\begin{table}[tbp]
    \caption{The hyperfine structure and hyperfine structure splitting of the 
             bound $S(L = 0)-$states of the $dd\mu$ ion (in $MHz$).}
      \begin{center}
      \begin{tabular}{llll}
        \hline\hline
         & $dd\mu$ & $(dd\mu)^{*}$\\
      \hline
$\epsilon_{J=\frac52}$ &  4.656669271$\cdot 10^6$ &  4.023227167$\cdot 10^6$ \\

$\epsilon_{J=\frac32}$ &  2.328339810$\cdot 10^6$ &  2.011617137$\cdot 10^6$ \\

$\epsilon_{J=\frac12}$ &  6.899579465$\cdot 10^0$ &  4.737788941$\cdot 10^0$ \\

$\epsilon_{J=\frac12}$ & -4.656669271$\cdot 10^6$ & -4.023227167$\cdot 10^6$ \\

$\epsilon_{J=\frac32}$ & -6.985012530$\cdot 10^6$ & -6.034846672$\cdot 10^6$ \\
      \hline
       \hline
$\Delta(\frac52 \rightarrow \frac32)$ & 2.328329461$\cdot 10^6$ & 2.01161003$\cdot 10^6$ \\

$\Delta(\frac32 \rightarrow \frac12)$ & 2.328332911$\cdot 10^6$ & 2.01161239$\cdot 10^6$ \\

$\Delta(\frac12 \rightarrow \frac12)$ & 4.656676170$\cdot 10^6$ & 4.02323191$\cdot 10^6$ \\

$\Delta(\frac12 \rightarrow \frac32)$ & 2.328343259$\cdot 10^6$ & 2.01161951$\cdot 10^6$ \\
       \hline\hline
   \end{tabular}
   \end{center}
   \end{table}

\end{document}